\documentclass[aip,jcp,reprint]{revtex4-2}
\usepackage[english]{babel}
\usepackage{amsmath,amsthm,latexsym,amssymb,amsfonts,epsfig}
\usepackage{bm}
\usepackage{dsfont}
\usepackage[left = 1.35cm, right = 1.35cm, top = 2.0cm, right = 2.0cm]{geometry}

\usepackage{graphicx}
\usepackage{color}

\definecolor{OliveGreen}{rgb}{0.1, 0.4, 0.1}
\definecolor{AppleGreen}{rgb}{0.55, 0.71, 0.0}

\usepackage{setspace}
\emergencystretch=\maxdimen
\usepackage{color}

\definecolor{OliveGreen}{rgb}{0.1, 0.4, 0.1}
\definecolor{caribbeangreen}{rgb}{0.0, 0.8, 0.6}
\definecolor{carminered}{rgb}{1.0, 0.0, 0.22}
    
\def\XXint#1#2#3{{\setbox0=\hbox{$#1{#2#3}{\int}$}
     \vcenter{\hbox{$#2#3$}}\kern-.5\wd0}}

\allowdisplaybreaks

\usepackage[colorlinks=true,citecolor=OliveGreen,linkcolor=OliveGreen,urlcolor=OliveGreen]{hyperref}

\usepackage{mathtools}
\usepackage{mathrsfs}

\usepackage{type1ec} %

\usepackage{amsbsy}

\newcommand{\bnabla}{\boldsymbol{\nabla}}
\newcommand{\ch}{\operatorname{ch}}
\newcommand{\sh}{\operatorname{sh}}
\renewcommand{\th}{\operatorname{th}} 

\newcommand{\sgn}{\operatorname{sgn}}

\begin{document}

\title{Elastic displacements and viscous hydrodynamic flows in wedge-shaped geometries with a straight edge: Green's functions for parallel forces}

\author{Abdallah Daddi-Moussa-Ider}
\email{admi2@open.ac.uk}
\thanks{Corresponding author.}
\affiliation{School of Mathematics and Statistics, The Open University, Walton Hall, Milton Keynes MK7 6AA, United Kingdom}

\author{Lukas Fischer}

\affiliation{Institut f\"ur Physik, Otto-von-Guericke-Universit\"at Magdeburg, Universit\"atsplatz 2, 39106 Magdeburg, Germany}

\author{Marc Pradas}

\affiliation{School of Mathematics and Statistics, The Open University, Walton Hall, Milton Keynes MK7 6AA, United Kingdom}

\author{Andreas M. Menzel}
\email{a.menzel@ovgu.de}

\affiliation{Institut f\"ur Physik, Otto-von-Guericke-Universit\"at Magdeburg, Universit\"atsplatz 2, 39106 Magdeburg, Germany}

\begin{abstract}
For homogeneous and isotropic linearly elastic solids and for incompressible fluids under low-Reynolds-number conditions the fundamental solutions of the associated continuum equations were derived a long time ago for bulk systems. That is, the corresponding Green's functions are available in infinitely extended systems, where boundaries do not play any role. However, introducing boundaries renders the situation significantly more complex. Here, we derive the corresponding Green's functions for a linearly elastic homogeneous and isotropic material in a wedge-shaped geometry. Two flat boundaries confine the material and meet at a straight edge. No-slip and free-slip conditions are considered. The force is oriented in a direction parallel to the straight edge of the wedge. Assuming incompressibility, our expressions also apply to the situation of low-Reynolds-number hydrodynamic viscous fluid flows. Thus, they may be used, for instance, to describe the motion of self-propelled objects guided by an edge or the distortion of soft elastic actuators in wedge-shaped environments of operation. 
\end{abstract}

\maketitle

\section{Introduction}
\label{sec:intro}

The basic equations of linear elasticity for an isotropic, spatially homogeneous elastic solid were outlined long ago in terms of the Navier-Cauchy equations~\cite{cauchy1828exercises}. Similarly, an isotropic, spatially homogeneous, incompressible viscous fluid under low-Reynolds-number conditions is described by the overdamped Stokes equations \cite{kim13}. In bulk, that is, in effectively infinitely extended systems, the fundamental solutions given by the Green's functions are available in terms of the well-known Kelvin tensor \cite{love2013treatise} and the Oseen tensor \cite{kim13}, respectively. They quantify the displacements and flows induced in the material by the localized application of a force at one point within the material. Due to the linearity of the underlying equations, the response to any bulk force density applied to the system is then obtained by straightforward convolution with the corresponding Green's function. 

In reality, boundaries limit the system size. Solving the equations mentioned above under limiting boundaries is a significantly more challenging task. Exact analytical expressions for associated Green's functions are available for half-space solutions that feature one flat, infinitely extended surface. From there, the system is assumed to extend into the complete infinite half-space on one side of the surface. Specific surface conditions are no-slip surfaces, which do not allow for any displacement nor flow at the boundary \cite{blake1971note, felderhof05, menzel2017force}, and free-slip conditions, which do not allow for any displacement nor flow normal to the surface but impose no restriction along the parallel direction \cite{menzel2017force}.
To some extent, the situations of sheets or films of finite thickness have also been addressed, where the system is bounded by two parallel flat surfaces but still infinitely extended to the sides \cite{liron76,felderhof06twoMem, felderhof10echoing, mathijssen2016hotspots, mathijssen2016hydrodynamics, daddi18jpcm, lutz2024internal}. Examples of three-dimensionally bounded geometries that are genuinely finite systems and have been repeatedly studied analytically are given by spherical systems \cite{walpole2002elastic, daddi2017hydrodynamic, daddi2017hydrodynamicII, daddi2018creeping, fischer2019magnetostriction, hoell2019creeping, sprenger2020towards, kree2021dynamics, fischer2020towards, fischer2024magnetic, kawakami2025migration}. The associated analytical expressions are complex. Still, the smoothness of the entire surface of the sphere favors analytical approaches. In other words, all the examples mentioned so far do not feature any kinks nor edges on their surfaces. 

In the present work, we turn away from these smooth geometries and address boundaries that contain one straight infinitely extended edge. The elastic or fluid material is confined to a wedge-like geometry with the edge forming the tip of this wedge. 
Flows at low Reynolds numbers in incompressible viscous fluids near three-dimensional corners were originally investigated by Sano and Hasimoto, who examined a series of progressively intricate wedge-shaped geometries~\cite{sano76, sano1977slow, sano1978effect, hasimoto80, sano1977slow_thesis}. More recently, the three-dimensional Stokes flow near a corner has been reexamined using techniques from complex analysis~\cite{dauparas2018leading, dauparas2018stokes}.
For overdamped viscous fluid flows under conserved volume, a corresponding Green's function has recently been derived. Yet, it was restricted to the case of free-slip surface conditions on the surfaces of the wedge and a $2\pi$-commensurate opening angle of the wedge \cite{sprenger2023microswimming}. Instead, here we generalize these considerations to linearly elastic, possibly compressible materials, coupled with no-slip and free-slip surface conditions, and allowing for arbitrary opening angles of the wedge. Our results equally apply to low-Reynolds-number flows of incompressible fluids by setting the material parameter related to compressibility in linearly elastic systems to the value that ensures incompressibility.  
Together with the solution obtained in Ref.~\onlinecite{daddi2025elastic} for the case of a perpendicular point force, this allows the construction of the general solution for an arbitrary force direction.

We proceed in Sec.~\ref{sec:math} by presenting the underlying continuum equations for linearly elastic materials. For incompressible fluid flows at low Reynolds numbers, we have to identify the shear modulus with the shear viscosity, elastic displacements with flow velocity, and set the Poisson ratio to one half. 
We continue in Sec.~\ref{sec:greens_function_FKL}, where the main derivations are presented together with the results for the Green's functions in the Fourier-Kontorovich-Lebedev space under different boundary conditions. 
We provide expressions of the Green's functions in Sec.~\ref{sec:greens_function_real} by applying the inverse transformation.  
In Sec.~\ref{sec:greens_function_planar_boundary}, we recover known results in the limit of a planar boundary, corresponding to an opening angle of the wedge of $\pi$.
Several technical details and parts of the calculations are shifted to the Appendices. 
Our conclusions are added in Sec.~\ref{sec:conclusions}.

\section{Governing equations}
\label{sec:math}

\begin{figure}
    \centering
    \includegraphics[width=1\linewidth]{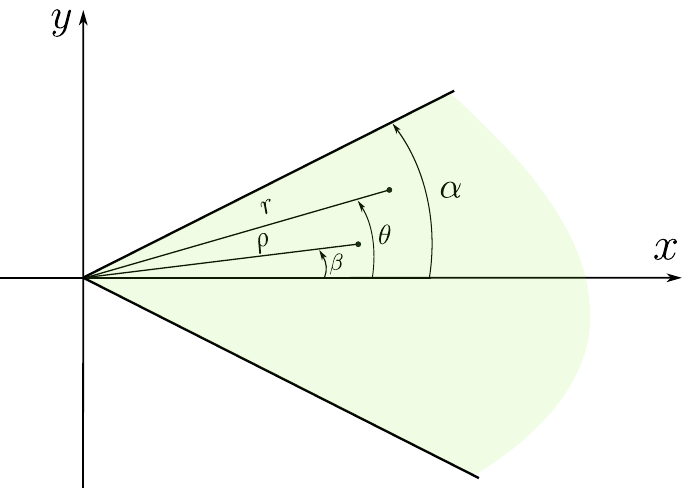}
    \caption{Schematic illustration of the setup.  
A wedge-shaped elastic medium with a straight edge aligned along the $z$-axis is confined by two planar boundaries and described in the cylindrical coordinate system $(r, \theta, z)$. The boundaries confining the elastic medium are located at $\theta = \pm \alpha$. A point-force singularity is applied along the $z$-direction and located at $(\rho, \beta, 0)$. The $z$-direction here points out of the plane towards the reader.}
    \label{fig:illustration}
\end{figure}

We consider a wedge-shaped elastic medium with a straight edge with the edge forming its tip. Figure~\ref{fig:illustration} graphically illustrates the system geometry. 
We use a cylindrical coordinate system $(r, \theta, z)$, with the straight edge of the wedge aligned along the $z$-axis. The boundaries of the elastic medium with the external medium are found at $\theta = \pm\alpha$, with $\alpha \in (0,\pi/2]$.
The special case $\alpha = \pi/2$ corresponds to a semi-infinite elastic medium with a planar surface. Thus, the opening angle of the wedge is given by the angle $2\alpha$. We assume that there is a force singularity in the cylindrical coordinate system located at $( r,\theta,z) = (\rho, \beta, 0)$, directed along the $z$-axis, perpendicular to the plane of the drawing in Fig.~\ref{fig:illustration}.

In the present paper, we examine three cases: (i) a no-slip  (NOS) boundary condition  on both walls located at $\theta = \pm\alpha$, (ii) a free-slip (FRS) boundary condition  on both walls located at $\theta = \pm\alpha$, and (iii) mixed boundary conditions with NOS at the wall located at $\theta = -\alpha$ and FRS at $\theta = \alpha$. 
The opposite case, with FRS at $\theta=-\alpha$ and NOS at $\theta=\alpha$, can be obtained by simply rotating the frame of reference around the $x$-axis by an angle of $\pi$.

\subsection{Solution methodology}

The Navier-Cauchy equation governs the displacement field $\bm{u}(\bm{x})$ of the material elements within an isotropic, homogeneous, linearly elastic medium. All transients of applying a constant force density are assumed to have decayed, so that the resulting static deformed state is described by
\begin{equation}
    \mu {\Delta} \bm{u}(\bm{x}) + \left( \lambda + \mu \right) \boldsymbol{\nabla} \left( \boldsymbol{\nabla} \cdot \bm{u}(\bm{x}) \right) + \bm{f}(\bm{x}) = \mathbf{0} \, . 
    \label{eq:Navier-Cauchy-Eqs}
\end{equation}
Here, $\lambda$ and $\mu$ represent the first and second Lamé parameters, respectively. $\bm{f}(\bm{x})$ denotes the force density field acting on the elastic material. The two coefficients are related to the Poisson ratio $\nu \in (-1,1/2]$ via
\begin{equation}
    \nu = \frac{\lambda}{2 \left( \lambda+\mu\right)} \, .
\end{equation}
A general solution to Eq.~\eqref{eq:Navier-Cauchy-Eqs} is given by
\begin{equation}
    \bm{u} = \bnabla \left( \bm{x} \cdot \boldsymbol{\Phi} + \Phi_w \right) - (\sigma+1) \boldsymbol{\Phi} \, , 
\end{equation}
where $\bm{x} = (x, y, z)$ represents the position vector in Cartesian coordinates and $\boldsymbol{\Phi} = \Phi_x \,\hat{\bm{e}}_x + \Phi_y \,\hat{\bm{e}}_y + \Phi_z \, \hat{\bm{e}}_z$. We omit writing the dependence of $\bm{u}$ and $\boldsymbol{\Phi}$ on $\bm{x}$ from this point onward.
Here, $\Phi_j$ are harmonic functions that satisfy Laplace's equation, $\Delta \Phi_j = 0$, for  $j \in \{x, y, z, w\}$.
In addition, we defined
\begin{equation}
    \sigma = 3-4\nu \,\in \, [1,7) \, .
\end{equation}
For $\nu = 1/2$, we obtain $\sigma = 1$, thereby recovering the classical form of solution proposed by Imai~\cite[p.~313]{imai1973ryutai}.

The problem can be solved more conveniently using a cylindrical coordinate system by expressing the displacement field in terms of its radial, azimuthal, and axial components. They are denoted as $u_r$, $u_\theta$, and~$u_z$, respectively. Applying the standard relations of coordinate transformation from Cartesian to cylindrical coordinates, the components of the displacement field in the cylindrical system are given by
\begin{subequations}
\label{eq:displacement_field}
    \begin{align}
    u_r &= -\sigma \, \Phi_r + r\, \frac{\partial \Phi_r}{\partial r} + z\, \frac{\partial \Phi_z}{\partial r} + \frac{\partial \Phi_w}{\partial r} \, , \\
    u_\theta &= \frac{\partial \Phi_r}{\partial \theta} + \frac{z}{r} \frac{\partial \Phi_z}{\partial \theta} + \frac{1}{r} \frac{\partial \Phi_w}{\partial \theta} - (\sigma+1) \Phi_\theta \, , \\
    u_z &= r\, \frac{\partial \Phi_r}{\partial z} -\sigma \, \Phi_z + z \, \frac{\partial \Phi_z}{\partial z} + \frac{\partial\Phi_w}{\partial z} \, ,
\end{align}
\end{subequations}
where the radial and azimuthal components of~$\boldsymbol{\Phi}$ are given, respectively, by
\begin{subequations} \label{eq:PhirPhiThe}
    \begin{align}
    \Phi_r &= \Phi_x \cos\theta + \Phi_y \sin\theta \, , \\
    \Phi_\theta &= \Phi_y \cos\theta - \Phi_x \sin\theta \, .
\end{align} 
\end{subequations}
We recall that the Laplacian of a function \( f \) in cylindrical coordinates is given by
\begin{equation}
    \Delta f = \frac{1}{r} \frac{\partial}{\partial r}
    \left( r \, \frac{ \partial f }{\partial r} \right)
    + \frac{1}{r^2} \frac{\partial^2 f}{\partial \theta^2}
    + \frac{\partial^2 f}{\partial z^2} \, .
    \label{eq:Laplcian}
\end{equation}
It follows from the coordinate transformation of the Laplace equation, Eq.~(\ref{eq:Laplcian}), that \(\Phi_r\) and \(\Phi_\theta\) satisfy
\begin{subequations} \label{eq:EqLapPhiRLapPhiThe}
    \begin{align}
    \Delta \Phi_r &= \frac{1}{r^2}
    \left( \Phi_r + 2 \, \frac{ \partial \Phi_\theta }{\partial \theta} \right) , \\
    \Delta \Phi_\theta &= \frac{1}{r^2}
    \left( \Phi_\theta - 2 \, \frac{ \partial \Phi_r }{\partial \theta} \right) ,
\end{align} 
\end{subequations}
where the terms on the right-hand side arise from the coupling between components in a curvilinear coordinate system.  
In an infinitely extended bulk linearly elastic medium, the Green's function corresponding to a point force applied at position $\bm{x}_0$ reads\cite{stakgold2011green}
\begin{equation}
    \mathcal{G}_{ij}^\infty = \frac{1}{4\pi \mu \left(\sigma +1\right)}
    \left( \frac{\sigma}{s} \, \delta_{ij} + \frac{ s_i s_j }{s^3} \right),
    \label{eq:green-infinity}
\end{equation}
where  $\bm{s} = \bm{x} - \bm{x}_0$ and \( s := |\bm{s}| \) denotes the distance from the applied force singularity.
Thus, for $\sigma = 1$, we recover the hydrodynamic Green's function, commonly known as the Oseen tensor~\cite{oseen28}, with $\mu$ representing the shear viscosity~\cite{happel12}.

In a wedge-shaped geometry, the general solution is expressed as
\begin{equation}
    \Phi_j = \phi_j^\infty + \phi_j \, ,
\end{equation}
where \(\phi_j^\infty\), with \( j \in \{r,\theta,z,w\} \), are the harmonic functions in an unbounded medium. The fields $\phi_j$ represent unknown functions to be determined, subject to the boundary conditions imposed at the boundaries of the wedge.

In the following, we consider combinations of two types of boundary conditions, namely no-slip and free-slip boundary conditions. 
A NOS boundary condition represents a condition where there is no relative motion between the material and the boundary, that is  $\bm{u} = \bm{0}$ on the boundary. 
A FRS boundary condition is characterized by both impermeability, where the normal component of the displacement satisfies $u_\theta = 0$,  and no-shear conditions, given by $\partial u_r / \partial \theta = \partial u_z / \partial \theta = 0$.

\subsection{Fourier-Kontorovich-Lebedev transform}

To determine the solution for the Green's function, we use the Fourier-Kontorovich-Lebedev (FKL) transform, a technique well-suited for solving elasticity and hydrodynamics problems involving wedge-shaped geometries. This method involves transforming the axial and radial coordinates into the axial and radial wavenumbers, denoted as $k$ and~${p}$, respectively.

On one hand, the Fourier transform is typically applied only with respect to the $z$-coordinate. Frequently, corresponding systems are unbounded in the $z$-direction and exhibit translational invariance, making it natural to use the Fourier transform to handle these infinite or semi-infinite domains. Consequently, differential operators in $z$ are transformed into algebraic ones, which simplifies the analysis.
On the other hand, the radial coordinate $r$ ranges from zero to infinity, which makes the Kontorovich–Lebedev (KL) transform well-suited for this variable. The KL transform is effective for radial problems involving wedge or conical geometries, because it diagonalizes operators that involve Bessel functions, which naturally arise in cylindrical coordinates.
Combining the two transforms allows for a separation of variables that aligns with the considered geometry and the associated boundary conditions. The Fourier transform reduces the dimensionality by handling the axial direction. Simultaneously, the KL transform efficiently manages the radial dependence in a manner consistent with the wedge boundaries and the spectral properties of the Laplacian in the radial coordinate. This combined approach leverages the strengths of each transform for the respective variable.

While the Fourier transform is widely known and its application is well-understood, the KL transform is much less familiar.
In the late 1930s, Kontorovich and Lebedev originally introduced these transforms to solve certain boundary value problems~\cite{kontorovich1938one, kontorovich1939method}. Later, further applications to boundary value problems were provided by Lebedev and Kontorovich~\cite{kontorovich1939application}, while the mathematical theory was developed by Lebedev~\cite{lebedev1946, lebedev1949}, see Erd{\'e}lyi \textit{et al.}~\cite[p.~75]{erdelyi1953higher} for further details on this transform. 
The KL transform has been extensively applied to a range of physical problems involving wedge geometries, particularly in the context of electromagnetic scattering and diffraction~\cite{lowndes1959application, rawlins1999diffraction, antipov2002diffraction, salem2006electromagnetic, hwang2009scattering, kim2009electromagnetic, shanin2011modified, eom2014integral, lyalinov2016integral}, as well as in the study of fluid flows~\cite{waechter1969steady, waechter1969steadyB}. 
Other studies have employed the KL transform in the presence of mixed boundary conditions~\cite{rao1981diffraction, naylor1966direct}.

We start by defining the forward Fourier transform with respect to \( z \) of a given function \( f(r,z) \) as
\begin{equation}
    \hat{f}(r, k) 
    := \mathscr{F} \left\{ f \right\}
    = 
    \int_{-\infty}^\infty f(r,z) \, e^{ikz} \, \mathrm{d} z \, , 
\end{equation}
where we have adopted the convention of a positive exponent. Next, we define the Kontorovich-Lebedev transform with respect to \( r \) as
\begin{equation}
    \widetilde{f}({p}, k) 
    := \mathscr{K}_{i{p}} \left\{ \hat{f} \right\}
    = \int_0^\infty \hat{f}(r, k) K_{i{p}} (|k|r) \, r^{-1} \, \mathrm{d} r  \, ,
\end{equation}
where \( K_{i{p}} (|k|r) \) denotes the modified Bessel function of the second kind~\cite{abramowitz72} of purely imaginary order \( i{p} \). Note that we use the hat to denote the Fourier transform of the original function, and the tilde to represent the Kontorovich-Lebedev transform of the Fourier-transformed function. 
The polar angle \( \theta \) remains unaffected by these transformations. 
Consequently, we transform the partial differential equation, Eq.~\eqref{eq:Navier-Cauchy-Eqs}, governing the displacement field in an elastic medium into a system of ordinary differential equations for the variable~\( \theta \).
Note that for a positive argument, \( K_{i{p}} \) yields a real number when \( {p} \) is constrained to be real.

In the following, we denote the hyperbolic sine, cosine, and tangent functions by sh, ch, and th, respectively.
Then, the inverse transforms are given by
\begin{equation}
    \hat{f}(r,k) = 
    \frac{2}{\pi^2} \int_0^\infty \widetilde{f}({p}, k) K_{i{p}} (|k|r) \sh (\pi{p}) \, {p} \, \mathrm{d} {p} \,  \label{eq:inv_KL}
\end{equation} 
for the inverse Kontorovich-Lebedev transform, and
\begin{equation}
    f(r,z) = \frac{1}{2\pi} 
    \int_{-\infty}^\infty \hat{f}(r,k) \, e^{-ikz} \, \mathrm{d} k \,  \label{eq:inv_F}
\end{equation}
for the inverse Fourier transform.
In many practical situations, performing the inverse Fourier transform is much easier. Therefore, in most cases, the solution will be expressed as a single infinite integral over the radial wavenumber ${p}$, which can be evaluated using numerical methods.

For convenience, we introduce the notation for the combined transform,
\begin{equation}
    \widetilde{f} = \mathscr{T}_{i{p}} \left\{ f \right\}
    = \mathscr{K}_{i{p}} \left\{ \mathscr{F} \left\{ f \right\} \right\} \, ,
\end{equation}
which we refer to as the FKL transform.
We show in Appendix~\ref{appendix:FKL_properties} that the FKL transform of $r^2 \Delta f$ is given by
\begin{equation}
    \mathscr{T}_{i{p}} \left\{ r^2 \Delta f \right\}
    = \left( \frac{\partial^2}{\partial\theta^2}-{p}^2 \right) \widetilde{f} \, .
\end{equation}
Consequently, the FKL transform of the Laplace equation yields a homogeneous second-order differential equation for the transformed function.
Throughout this paper, a prime will denote the derivative with respect to the polar angle \( \theta \), and we introduce, for convenience, the notation \( \mathscr{T}_{i{p}}^{\,\prime} \{ f \} = \partial \widetilde{f} / \partial \theta \).

\section{Green's function in FKL space}
\label{sec:greens_function_FKL}

\subsection{Free-space Green's function}

In an unbounded medium, the displacement field induced by a point force applied in the \( z \)-direction follows from
\begin{equation}
    \phi_r^\infty = \phi_\theta^\infty = \phi_w^\infty = 0 \, , \quad
    \phi_z^\infty = \chi \, ,  
\end{equation}
where
\begin{equation}
    \chi = -\frac{q_\parallel}{s} \, , \quad
    q_\parallel = \frac{F_z}{4\pi \mu \left(1+\sigma \right)} \, , \label{eq:chi}
\end{equation}
with
\begin{equation}
    s = \left( r^2+\rho^2-2\rho r \cos (\theta-\beta)+z^2 \right)^\frac{1}{2} \, .
    \label{eq:s}
\end{equation}
$s$ represents the distance from the position of the force singularity in cylindrical coordinates. Again, the force singularity is assumed to be located at $( r,\theta,z) = (\rho, \beta, 0)$.

Using the parity of~\( \chi \) as defined by Eq.~\eqref{eq:chi} with respect to $z$ in conjunction with the integral that defines the zeroth-order modified Bessel function of the second kind, see Watson~\cite[p.~183]{watson22},
\begin{equation}
    K_0(x) = \int_{0}^\infty 
    \frac{\cos(xt)\, \mathrm{d}t}{\left( t^2+1\right)^\frac{1}{2}} \qquad (x>0) \, , 
\end{equation}
the Fourier transform of \( \chi \) is obtained as
\begin{equation}
    \hat{\chi} = \mathscr{F} \left\{ \chi \right\}
    = -2q_\parallel K_0(|k|R) \, , \label{eq:chi_hat}
\end{equation}
where 
\begin{equation}
    R = s(z=0) = \left( r^2+\rho^2-2\rho r\cos(\theta-\beta)\right)^\frac{1}{2} \, .
\end{equation}
We utilize the Table of Integral Transforms by Erdélyi \textit{et al.}~\cite[p.~175]{erdelyi54} to derive the FKL transform of Eq.~\eqref{eq:chi_hat} as
\begin{equation}
    \widetilde{\chi} =
    -2\pi q_\parallel \, 
    \frac{\ch \left( ( \pi-|\theta-\beta| ){p} \right)}{{p} \sh(\pi{p})} \, K_{i{p}} (|k|\rho) \, .
    \label{eq:chi_FKL}
\end{equation}

\subsection{General form of the solution}

Since $\Phi_j$, with $j \in \{x, y, z, w\}$, are harmonic functions, it follows that their corresponding FKL transforms satisfy the second-order ordinary differential equations 
\begin{equation}
    \widetilde{\phi}_j^{\,\prime\prime}-{p}^2 \,\widetilde{\phi}_j = 0 \, , \label{eq:FKL-laplace}
\end{equation}
for $j \in \{x, y, z, w\}$.
Therefore, a general solution to Eq.~\eqref{eq:FKL-laplace} is given by
\begin{equation}
    \widetilde{\phi}_j = A_j \sh (\theta{p}) + B_j \ch (\theta{p}) \, . \label{eq:solution_form}
\end{equation}
Here, $A_j$ and~$B_j$ are eight coefficients to be determined from the boundary conditions.
It is important to note that $\widetilde{\phi}_r$ and $\widetilde{\phi}_\theta$ are not harmonic functions, see Eq.~\eqref{eq:EqLapPhiRLapPhiThe}. Thus, to them the form of solutions in Eq.~\eqref{eq:solution_form} does not apply.

Counting all components, there are six equations resulting from the boundary conditions imposed on both surfaces. In contrast to this, there are eight unknowns $A_j$ and~$B_j$, $j \in \{x, y, z, w\}$. Thus, we have the flexibility to choose two of the eight unknowns. We select the ones that yield the simplest possible form. 

In this context, we remark that in three-dimensional Stokes flow, the solution is commonly described using three independent harmonic functions, known as Lamb's general solution~\cite{happel12}: the harmonic velocity potential, the harmonic stream function (vector potential), and the harmonic pressure field. 
The same applies to linear elasticity, where the evolving field is displacement instead of flow velocity.
Since Imai’s solution~\cite{imai1973ryutai} involves four harmonic functions, one is expected to be linearly dependent on the others.
This would imply that, no matter which choices are made, one always gets the same physical solution upon back-transform. In other words, the choices do not affect the solutions in real space.

Before proceeding, we first state the following identity, which we demonstrate in  Appendix~\ref{appendix:FKL_properties},
\begin{equation}
    \mathscr{T}_{i{p}} \left\{ \frac{z}{r} \, f \right\} 
    =  \mathcal{C}_2^+
    + \mathcal{C}_2^- , \label{eq:FKL_rzf}
\end{equation}
with the abbreviations
\begin{equation}
    \mathcal{C}_2^\pm = 
    -\frac{\sgn k}{2{p}} \left(
    i{p} \pm 1 \pm k\, \frac{\partial}{\partial k} \right) \mathscr{T}_{i{p}\pm 1} \left\{ f\right\} ,
\end{equation}
where $\sgn x = x/|x|$ denotes the sign function.

For the following analytical consideration, we express the unknown coefficients $A_j$ and $B_j$, $j\in\{x,y,z\}$, in specific forms. 
The components in the axial direction are expressed as
\begin{subequations}
    \begin{align}
    A_z &= \frac{2\pi}{{p}\sh(\pi{p})} \, q_\parallel
\Lambda_z K_{i{p}}(|k|\rho) \, , \\
    B_z &= \frac{2\pi}{{p}\sh(\pi{p})} \, q_\parallel
\Lambda_z^\dagger K_{i{p}}(|k|\rho) \, .
\end{align}
\end{subequations}
Similarly, we cast the components in the radial-azimuthal plane as
\begin{subequations}
    \begin{align}
    A_j &= \frac{8i\pi}{{p}\sh(\pi {p})} \, q_\parallel \rho k \, \Lambda_j K_{i{p}} ( |k|\rho ) \,
     \, , \\
    B_j &= \frac{8i\pi}{{p}\sh(\pi {p})} \, q_\parallel \rho k \, \Lambda_j^\dagger K_{i{p}} ( |k|\rho ) \, ,
\end{align}
\end{subequations}
for $j \in \{x,y\}$.
Here, $\Lambda_j$ and $\Lambda_j^\dagger$ for $j \in \{x,y,z\}$ are functions of ${p}, \alpha, \beta,$ and $\sigma$ only and depend on the type of boundary conditions prescribed at $\theta = \pm\alpha$. Their expressions are provided below for the combinations of 2NOS, 2FRS, and NOS--FRS boundary conditions.

\subsection{No-slip--no-slip boundary conditions (2NOS)}

Requiring that the displacement field vanishes completely at both surfaces, the boundary conditions follow as
\begin{subequations}
	\begin{align}
	 \left( r\, \frac{\partial }{\partial r} -\sigma \right) \phi_r + z\, \frac{\partial}{\partial r} \left( \phi_z + \chi \right) + \frac{\partial \phi_w}{\partial r}
         &= 0 \, 
        \, , \\
	\frac{\partial \phi_r}{\partial \theta} + \frac{z}{r} \frac{\partial }{\partial \theta} \left( \phi_z + \chi \right) + \frac{1}{r} \frac{\partial \phi_w}{\partial \theta} - (\sigma+1) \phi_\theta  &= 0 \, , \\
	r\, \frac{\partial \phi_r}{\partial z} +  \left( z \, \frac{\partial }{\partial z} -\sigma \right) \left( \phi_z+\chi \right) + \frac{\partial\phi_w}{\partial z} &= 0 \, ,
	\end{align}
\end{subequations}
which are evaluated at $\theta = \pm \alpha$.
Therefore, the simplest choice satisfying the no-slip boundary conditions is
\begin{subequations} 
    \label{eq:BCsNS}
    \begin{align}
    \phi_r = 0 \, , \qquad \phi_w = 0 \, , \qquad \phi_z  +\chi &= 0 \, , \\[5pt]
    \frac{\partial \phi_r}{\partial \theta} - (\sigma+1) \,\phi_\theta 
    + \frac{1}{r} \frac{\partial \phi_w}{\partial \theta}
    + \frac{z}{r} \frac{\partial \phi_z}{\partial \theta}
    +\frac{z}{r} \frac{\partial \chi}{\partial \theta} &= 0 \, ,  
\end{align}
\end{subequations}
imposed at $\theta = \pm \alpha$.
We could have chosen alternative representations, but given the flexibility in selecting harmonic functions as discussed above, we opted for the form of solution that we consider simplest.

The FKL transform of the 2NOS boundary conditions, stated by Eqs.~\eqref{eq:BCsNS}, is given by
\begin{subequations}
    \begin{align}
    \widetilde{\phi}_r = 0 \, , \qquad \widetilde{\phi}_w = 0 \, , \qquad \widetilde{\phi}_z  + \widetilde{\chi} &= 0 \, , \\[5pt]
    \widetilde{\phi}_r^{\,\prime} - (\sigma+1) \,\widetilde{\phi}_\theta 
    + \mathscr{T}_{i{p}}^{\,\prime} 
    \left\{ \frac{\phi_w}{r} +\frac{z}{r} \left( \phi_z + \chi \right) \right\} &= 0 \, ,  
\end{align}
\end{subequations}
evaluated at $\theta = \pm \alpha$.
We note that \( \widetilde{\phi}_r \) and \( \widetilde{\phi}_\theta \) are derived from Eqs.~\eqref{eq:PhirPhiThe}, and the derivative of $\widetilde{\phi}_r$ with respect to~$\theta$ is given by
\begin{equation}
    \widetilde{\phi}_r^{\,\prime} = 
    \left( \widetilde{\phi}_y + \widetilde{\phi}_x^{\,\prime} \right) \cos\theta
    + \left( \widetilde{\phi}_y^{\,\prime} - \widetilde{\phi}_x \right) \sin\theta \, .
\end{equation}
Since \( \widetilde{\phi}_w \) vanishes at \( \theta = \pm \alpha \), it follows from the general solution given by Eq.~\eqref{eq:solution_form} that \( \widetilde{\phi}_w = 0 \) everywhere.
However, this is not the case for $\widetilde{\phi}_r$, because Eq.~\eqref{eq:solution_form} does not apply to $j = r$.
Thus, the boundary conditions simplify to
\begin{subequations}  \label{eq:BC_NS}
    \begin{align}
    \widetilde{\phi}_r = 0 \, , \qquad
    \widetilde{\phi}_z  + \widetilde{\chi} &= 0 \, , \label{eq:BC_NS_1} \\[5pt]
    \widetilde{\phi}_r^{\,\prime} - (\sigma+1) \, \widetilde{\phi}_\theta 
    + \mathscr{T}_{i{p}}^{\,\prime} \left\{ \frac{z}{r} \left( \phi_z + \chi \right) \right\} &= 0  \, ,  \label{eq:BC_NS_2} 
\end{align}
\end{subequations}
imposed at $\theta = \pm\alpha$.

After simplification, it follows from Eqs.~\eqref{eq:chi_FKL} and \eqref{eq:FKL_rzf} that for the free-space contribution we obtain
\begin{equation}
    \mathscr{T}_{i{p}}^{\,\prime}
    \left\{ \frac{z}{r}\, \chi \right\}
    = ik\rho \, \widetilde{\chi} \, \sin(\beta-\theta) \, .
\end{equation}

Using the boundary condition for $\widetilde{\phi}_z$, that is, the second equation in Eqs.~\eqref{eq:BC_NS_1}, we obtain
\begin{subequations}
    \begin{align}
    \Lambda_z &= \frac{\sh(\beta{p}) \sh ( (\pi-\alpha){p})}{\sh (\alpha{p})} \, , \\
    \Lambda_z^\dagger &= \frac{\ch(\beta{p}) \ch ( (\pi-\alpha){p})}{\ch (\alpha{p})} \, . 
\end{align}
\end{subequations}
Based on the boundary conditions for \( \widetilde{\phi}_x \) and \( \widetilde{\phi}_y \), specifically the first equation in Eqs.~\eqref{eq:BC_NS_1} and \eqref{eq:BC_NS_2}, the solution for the four remaining coefficients is obtained by solving a linear system of four unknowns, given by
\begin{equation}
    \Lambda_x = h_1 \sin\alpha \ch (\alpha{p})  \, , \quad
    \Lambda_y =  h_2 \cos\alpha \ch (\alpha{p}) \, ,
\end{equation}
together with
\begin{equation}
         \Lambda_x^\dagger = -h_2 \sin\alpha\sh(\alpha{p})  \, , \quad
    \Lambda_y^\dagger = -h_1 \cos\alpha\sh(\alpha{p})  \, ,
\end{equation}
where \( h_1 \) and \( h_2 \) are given by
\begin{subequations}
    \begin{align}
    h_1 &= \sh(\pi{p}) \big( \cos\alpha\sin\beta\sh(\alpha{p})\ch(\beta{p}) \notag \\
    &\quad-\sin\alpha\cos\beta\ch(\alpha{p})\sh(\beta{p}) \big) / \Delta_+ \, ,\\[3pt]
    h_2 &= \sh(\pi{p}) \big( \sin\alpha\cos\beta\sh(\alpha{p})\ch(\beta{p}) \notag \\
    &\quad-\cos\alpha\sin\beta\ch(\alpha{p})\sh(\beta{p}) \big) / \Delta_- \, ,
\end{align}
\end{subequations}
with the denominators
\begin{equation}
    \hspace{-0.1cm}
    \Delta_\pm = \left( \sigma \sh(2\alpha{p})\pm {p} \sin(2\alpha) \right)
    \left( \ch (2\alpha{p}) \mp \cos(2\alpha) \right) .
\end{equation}
It is important to note that the coefficients are related by $B_x = -A_y \tan\alpha \th(\alpha{p})$ and $B_y = -A_x \cot\alpha \th(\alpha{p})$.

For \( \sigma = 1 \), corresponding to the hydrodynamic or incompressible limit, our results are fully consistent with those of Sano and Hasimoto~\cite{sano1978effect} and Dauparas and Lauga~\cite{dauparas2018leading}.
Here, we present not only the general solution for arbitrary \( \sigma \) but also additional details of the derivation and a clearer formulation of our results.

\subsection{Free-slip--free-slip boundary conditions (2FRS)}

Next, we impose the free-slip boundary condition on both surfaces by requiring that
\begin{equation}
    u_\theta = 0 \, , \quad
    \frac{\partial u_r}{\partial \theta} = 0 \, , \quad
    \frac{\partial u_z}{\partial \theta} = 0 \, 
\end{equation}
at $\theta = \pm \alpha$.
This implies that
\begin{equation}
    \frac{\partial \phi_r}{\partial \theta} = 0 \, , \,\,
    \phi_\theta = 0 \, , \,\,
    \frac{\partial \phi_w}{\partial \theta} = 0 \, , \,\,
    \frac{\partial}{\partial\theta}
    \left( \phi_z + \chi \right) = 0 \, , 
\end{equation}
evaluated at $\theta = \pm \alpha$.

From the form of the solution given by Eq.~\eqref{eq:solution_form}, it follows that $\phi_j = 0$ for $j \in \{x,y,w\}$, implying $A_j = B_j = 0$ for $j \in \{x,y,w\}$. Consequently, the only nonzero harmonic function is $\phi_z$, leaving us with the determination of $A_z$ and~$B_z$ alone.

In FKL space, the resulting equation for the boundary conditions reads
\begin{equation}
    \mathscr{T}_{i{p}}^{\,\prime} (\phi_z + \chi) = 0  \, , \label{eq:BC_FS}
\end{equation}
at $\theta = \pm\alpha$.
The corresponding solution is obtained as
\begin{subequations}
    \begin{align}
    \Lambda_z &= -\frac{\sh(\beta{p}) \ch ((\pi-\alpha){p})}{\ch(\alpha{p})} \, , \\
    \Lambda_z^\dagger &= -\frac{\ch(\beta{p}) \sh ((\pi-\alpha){p})}{\sh(\alpha{p})} \, .
\end{align}
\end{subequations}

\subsection{No-slip--free-slip boundary condition (NOS--FRS)}

Finally, we examine the case of mixed boundary conditions by imposing a NOS condition at $\theta = -\alpha$ and a FRS condition at $\theta = \alpha$.
Specifically, we require  Eqs.~\eqref{eq:BC_NS} to be satisfied at $\theta = -\alpha$ and Eq.~\eqref{eq:BC_FS} at $\theta = \alpha$. After simplification, we obtain for the axial components 
\begin{subequations}
    \begin{align}
    \Lambda_z &= -\frac{ \sh(\beta{p}) \ch((\pi-2\alpha){p}) + \ch(\beta{p})\sh(\pi{p})}{\ch(2\alpha{p})} \, , \\
    \Lambda_z^\dagger &= \frac{\ch(\beta{p}) \ch((\pi-2\alpha){p}) - \sh(\beta{p})\sh(\pi{p})}{\ch(2\alpha{p})} \, ,
\end{align}
\end{subequations}
while the components in the radial-azimuthal plane are
\begin{subequations}
    \begin{align}
    \Lambda_x = \Sigma_+ \sin\alpha\ch(\alpha{p}) \, , \quad
    \Lambda_y = \Sigma_- \cos\alpha\ch(\alpha{p}) \, ,
\end{align}
\end{subequations}
and
\begin{equation}
    \Lambda_x^\dagger = \Sigma_+^\dagger \sin\alpha\sh(\alpha{p}) \, , \quad
    \Lambda_y^\dagger = \Sigma_-^\dagger  \cos\alpha\sh(\alpha{p}) \, . 
\end{equation}
In these expressions, we have defined
\begin{subequations}
    \begin{align}
    \Sigma_\pm &= H \left( 1-\ch(2\alpha{p}) \pm \cos(2\alpha) \right) \, , \\
     \Sigma_\pm^\dagger &= H \left( 1+\ch(2\alpha{p}) \pm \cos(2\alpha) \right) \, ,
\end{align}
\end{subequations}
where
\begin{equation}
    H = \frac{\varphi(\alpha+\beta, \beta-3\alpha) \sh(\pi{p})}{\left( \sigma \sh(4\alpha{p})-{p}\sin(4\alpha) \right) \left( \ch(4\alpha{p})+\cos(4\alpha) \right)} \, ,
\end{equation}
with
\begin{equation}
    \varphi(a,b) = \sin(a) \sh(b{p}) + \sin(b) \sh(a{p}) \, .
\end{equation}
In the next section, we apply the inverse transforms to obtain integral expressions for the displacement field in real space.

\section{Green's function in real space}
\label{sec:greens_function_real}

What remains is the inverse FKL transform, which is based on Eqs.~\eqref{eq:inv_KL} and~\eqref{eq:inv_F}, defined by the double integral
\begin{equation}
    \hspace{-0.1cm}
    \phi_j = \frac{1}{\pi^3} \int_{-\infty}^\infty \mathrm{d}k \, e^{-ikz} 
    \int_0^\infty \widetilde{\phi}_j \, K_{i{p}}(|k|r) \sh(\pi{p}) {p} \, \mathrm{d}{p} \, .
\end{equation}
Fortunately, analytical progress can be made by first evaluating the integration with respect to \( k \).
For the axial component, we obtain
\begin{equation}
    \phi_z(r,\theta,z) =
    \int_{0}^\infty \mathcal{K}_{i{p}} (r,z) \, \psi_z(\theta, {p}) \, \mathrm{d}{p} \, , 
    \label{eq:phiZ_final}
\end{equation}
where
\begin{align}
    \mathcal{K}_{i{p}} (r,z) = \left(\tfrac{2}{\pi}\right)^2  
    \int_{0}^\infty \cos(kz) K_{i{p}}(k\rho) K_{i{p}} (kr) \, \mathrm{d}k  \, , \label{eq:Kinu_def}
\end{align}
and
\begin{align}
    \psi_z (\theta, {p}) &= q_\parallel \left(
\Lambda_z \sh(\theta{p}) + \Lambda_z^\dagger \ch(\theta{p}) \right) . \label{eq:FZ}
\end{align}
The improper integral defined by Eq.~\eqref{eq:Kinu_def} is convergent and its value can be found in classical textbooks as
\begin{equation}
    \mathcal{K}_{i{p}} (r,z) =  
    \frac{P_{i{p}-\frac{1}{2}} (\xi)}{\left( \rho r \right)^\frac{1}{2} \ch (\pi{p})} \, ,
    \label{eq:Kinu}
\end{equation}
see Gradshteyn and Ryzhik~\cite[p.~719]{gradshteyn2014table} Eq.~ET I 50(51) or the treatise by Prudnikov \textit{et al.}~\cite[p.~390]{prudnikov1992integrals} Eq.~2.16.36~(2).
Here, \( P_n \) denotes the Legendre function of the first kind of degree \( n \), with argument
\begin{equation}
    \xi = \frac{\rho^2+r^2+z^2}{2\rho r} .
\end{equation}
Similarly, we follow the same approach for $\phi_x$ and $\phi_y$ to obtain
\begin{equation}
    \phi_j (r,\theta, z) = \int_0^\infty \mathcal{Q}_{i{p}} (r,z) \, \psi_j(\theta, {p}) \, \mathrm{d}{p} \,  
    \label{eq:phiXY_final}
\end{equation}
for $j \in \{x,y\}$, where 
\begin{align}
    \hspace{-0.1cm}
    \mathcal{Q}_{i{p}}(r,z) = 
    \left( \tfrac{4}{\pi} \right)^2 
    \int_{0}^\infty \rho k\sin(kz) K_{i{p}}(k\rho) K_{i{p}} (kr) \, \mathrm{d}k  , \label{eq:Qinu_def}
\end{align}
and 
\begin{equation}
    \psi_j (\theta, {p})  = q_\parallel \left( \Lambda_j \sh(\theta{p}) + \Lambda_j^\dagger \ch(\theta{p}) \right) 
    \label{eq:FXFY}
\end{equation}
for $j\in \{x,y\}$.
Likewise, the improper integral given by Eq.~\eqref{eq:Qinu_def} can be evaluated analytically as
\begin{equation}
\mathcal{Q}_{i{p}} (r,z) = 
    \frac{z}{r} 
    \frac{P_{i{p} - \frac{1}{2}}^{-1} (\xi) }{ \left( \rho r\right)^\frac12 \left( \xi^2-1\right)^{\frac{1}{2}} }  \frac{4{p}^2+1}{\ch (\pi {p})} \, , \label{eq:Qnui}
\end{equation}
see Gradshteyn and Ryzhik~\cite[p.~727]{gradshteyn2014table} Eq.~ET I 107(61) or
Prudnikov \textit{et al.}~\cite[p.~389]{prudnikov1992integrals} Eq.~2.16.36~(1).
Here, \( P_n^m \) denotes the associated Legendre function of the first kind of degree \( n \) and order \( m \).

For clarity, we note that various definitions of \( P_n^m \) exist in the literature and in commercial computer algebra systems. Here, we adopt the definition
\begin{equation}
    P_n^m (\xi) = \left( \frac{\xi+1}{\xi-1} \right)^\frac{m}{2}
    {}_2\widetilde{F}_1 \left( -n,n+1; 1-m; \tfrac{1-\xi}{2} \right) , \label{eq:LegendreAssoc}
\end{equation}
with ${}_2\widetilde{F}_1$ representing the regularized hypergeometric function, which is related to the Gaussian or ordinary hypergeometric function by the expression ${}_2\widetilde{F}_1(a,b; c; z) = {}_2F_1(a,b; c; z)/\Gamma(c)$, with \( \Gamma \) denoting the Euler gamma function~\cite{abramowitz72}.
This is the definition implemented in Maple~\cite{maple24}. 
In Mathematica~\cite{Mathematica}, this is referred to as the associated Legendre function of the first kind of type~3.
It differs from the standard definition in Mathematica or the one referred to as type 2, which should be multiplied by \( i^m \) to obtain the desired result.
We note that both \( \mathcal{K}_{i{p}} \) and \( \mathcal{Q}_{i{p}} \) yield real values. Far from the singularity, $\mathcal{K}_{ip}$ and $\mathcal{Q}_{ip}$, along with their derivatives, vanish, thereby ensuring that the solution satisfies the regularity conditions at infinity.

\begin{figure*}
    \centering
    \includegraphics[scale=0.95]{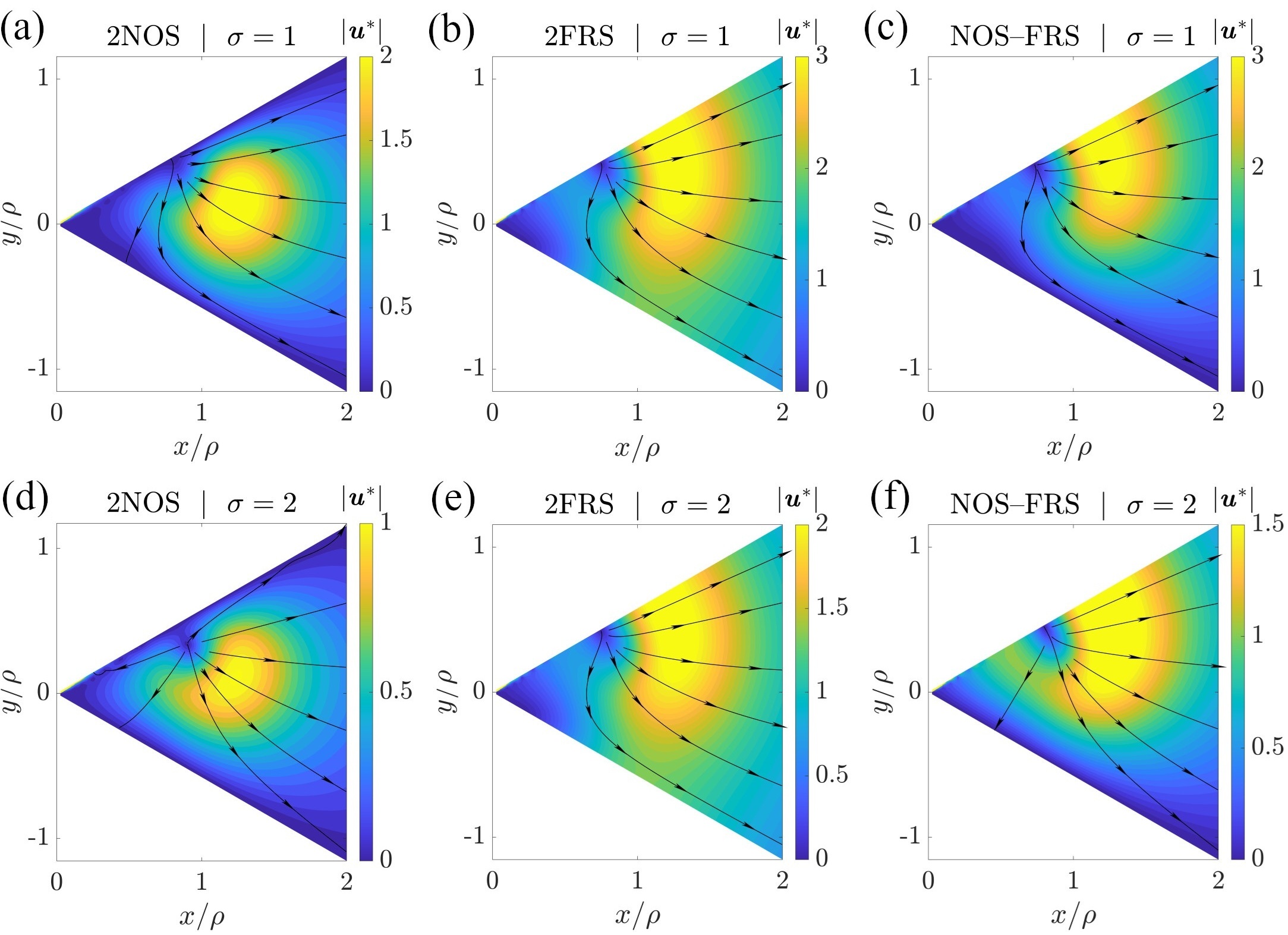}
    \caption{
Scaled displacement field $\bm{u}^* = \bm{u} / \left( F_z/\left(16\pi\mu\rho\right) \right)$ in the radial-azimuthal plane at $z/\rho = 1/2$ induced by a point force acting along the $z$-direction. A semi-opening angle of the wedge of $\alpha = \pi/6$ is considered, with the singularity positioned at $\beta = \alpha/2$. 
Panels shown in the top row (a)–(c) correspond to $\sigma = 1$, and those in the bottom row (d)–(f) to $\sigma = 2$. We address the different boundary conditions (a,d) 2NOS,  (b,e) 2FRS, and (c,f) NOS–FRS.
}
    \label{fig:displacement_field_R_The_plane}
\end{figure*}
To summarize, we have determined the harmonic functions and expressed the results as integrals over the radial wavenumber \( {p} \), see Eqs.~\eqref{eq:phiZ_final} and \eqref{eq:phiXY_final}. The improper integrals contained in these expressions could potentially be evaluated analytically using the method of residues. However, in this study, we opt for numerical integration to evaluate them, for simplicity. The infinite integrals can be transformed into well-behaved finite integrals over the domain \([0, 1]\), which are amenable to standard numerical integration methods, by introducing the change of variable \( u = {p}/({p} + 1) \), resulting in \( \mathrm{d}{p} = \mathrm{d} u/(1-u)^2 \).

The displacement field is then determined immediately from Eqs.~\eqref{eq:displacement_field} as
\begin{align}
    u_j (\bm{x}) &= u_j^\infty (\bm{x}) + \int_0^\infty U_j (\bm{x},{p}) \, \mathrm{d}{p} \, ,  
\end{align}
for $j \in \{r, \theta, z\}$, 
where
\begin{subequations} \label{eq:displacement_field_KQ}
    \begin{align}
    U_r &= \psi_r \left( r\, \frac{\partial }{\partial r} - \sigma \right) \mathcal{Q}_{i{p}} 
    + z \psi_z\, \frac{\partial \mathcal{K}_{i{p}}}{\partial r}  \, , \\
    U_\theta &= \left( \frac{\partial \psi_r}{\partial\theta} - (\sigma+1) \psi_\theta \right) \mathcal{Q}_{i{p}} + \frac{z}{r} \frac{\partial \psi_z}{\partial \theta} \, \mathcal{K}_{i{p}} \, , \\
    U_z &=  r \psi_r \, \frac{\partial \mathcal{Q}_{i{p}}}{\partial z} + \psi_z \left( z\, \frac{\partial}{\partial z} -\sigma \right) \mathcal{K}_{i{p}} \, ,
\end{align}
\end{subequations}
with $\psi_r = \psi_x\cos\theta+\psi_y\sin\theta$ and $\psi_\theta=\psi_y \cos\theta-\psi_x\sin\theta$.

\begin{figure*}
    \centering
     \includegraphics[scale=0.95]{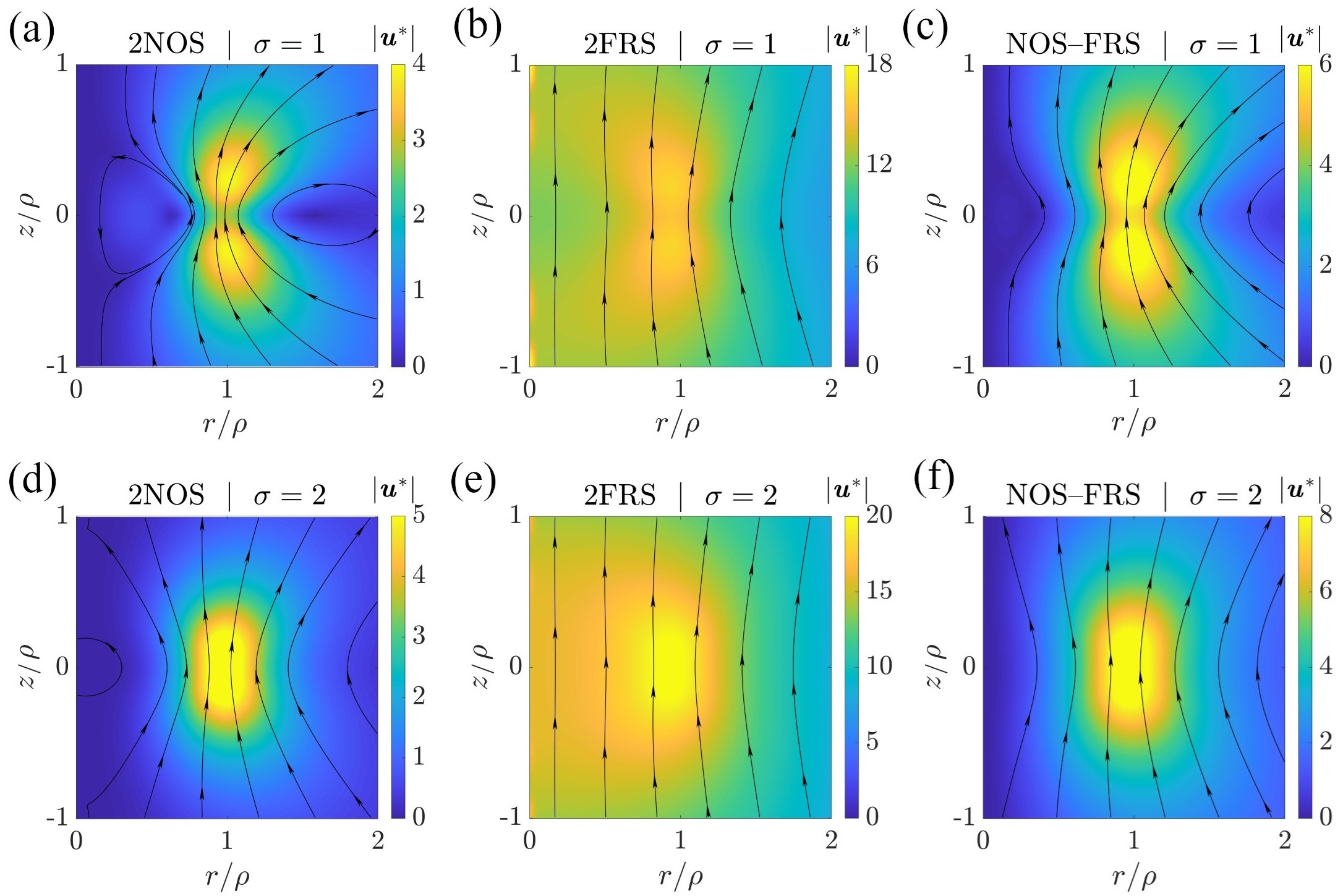}
    \caption{
Displacement field in the radial-axial plane of $\theta=0$ induced by a point force acting along the $z$-direction.
The parameters are identical to those in Fig.~\ref{fig:displacement_field_R_The_plane}.
Again, the scaled displacement field is defined as $\bm{u}^* = \bm{u} / \left( F_z/\left(16\pi\mu\rho\right) \right)$.
The top-row panels (a)–(c) correspond to $\sigma = 1$, while the bottom-row panels (d)–(f) correspond to $\sigma = 2$. Considered boundary conditions are (a,d) 2NOS, (b,c) 2FRS, and (c,f) NOS–FRS.
}
    \label{fig:displacement_field_R_Z_plane}
\end{figure*}

Since the derivatives of \( \mathcal{K}_{i{p}} \) and~\( \mathcal{Q}_{i{p}} \) with respect to~\( r \) and~\( z \) are required, their corresponding expressions are provided in  Appendix~\ref{appendix:derivatives}. 
It can readily be observed that $\partial \mathcal{K}_{i{p}}/\partial z = -\mathcal{Q}_{i{p}}/\left(4\rho\right)$, see the definitions of the integrals given by Eqs.~\eqref{eq:Kinu_def} and~\eqref{eq:Qinu_def}.

To verify our solutions, we tested them across a wide range of system parameter values. Depending on the boundary conditions applied, we either evaluate the displacements or the displacement gradients to ensure that the prescribed boundary conditions are met.

In this context, we remark an important aspect regarding the validity of the presented solutions. 
While the solutions for the 2NOS and 2FRS cases hold for the entire range $\alpha \in (0,\pi/2]$ for the semi-opening angle of the wedge, the solution for the NOS--FRS case is found to be valid only for $\alpha \in (0,\pi/4]$. For values beyond this range, although the equations remain satisfied in the FKL space, the boundary condition requiring vanishing azimuthal displacement on the NOS boundary is not met. This issue may stem from the fact that the integrals defining the FKL transform might not be well-defined in this specific range. Further investigation into this aspect would be valuable. 

\begin{figure*}
    \centering
     \includegraphics[scale=0.95]{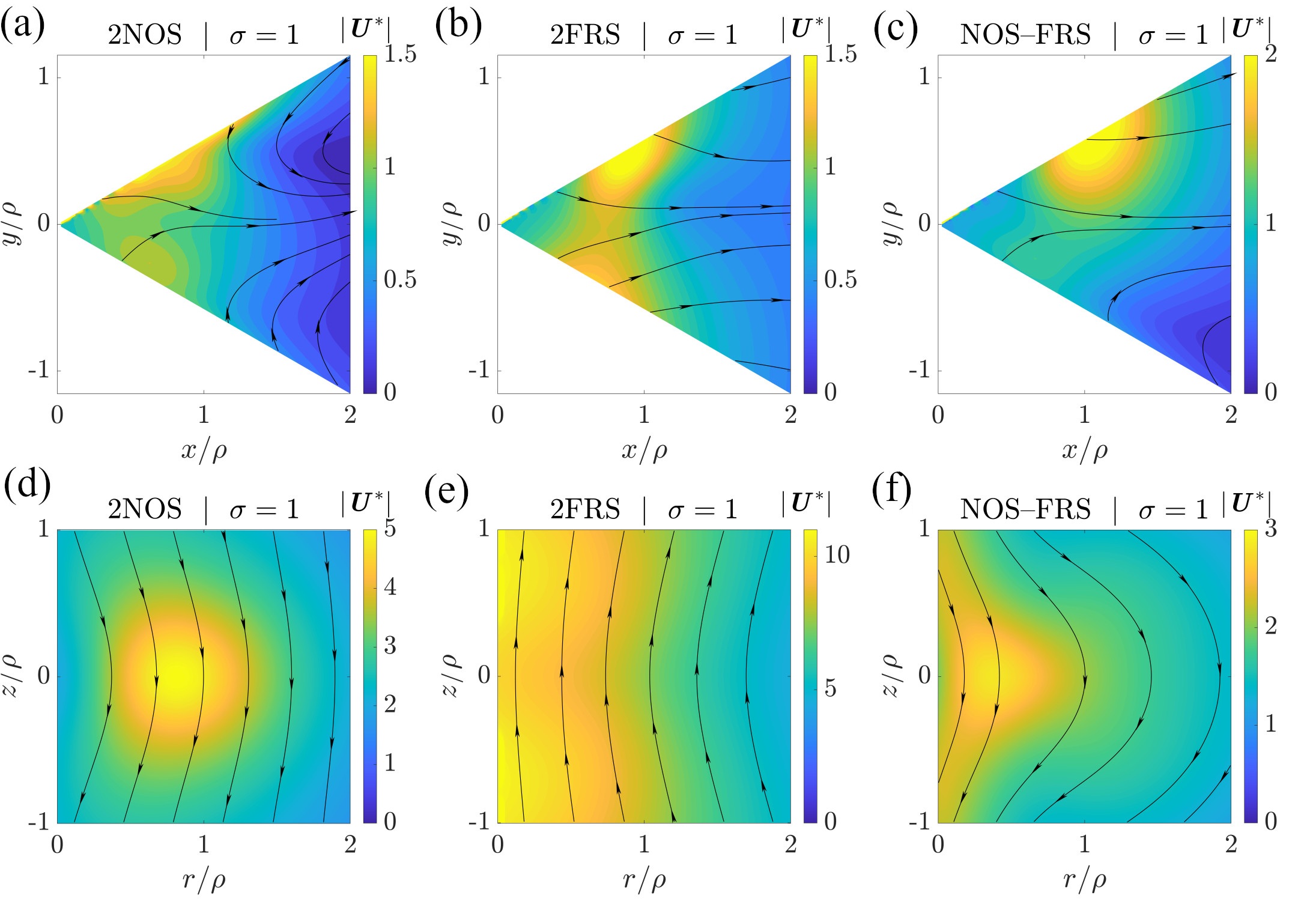}
    \caption{
Displacement field (a)--(c) in the radial–azimuthal plane at $z/\rho = 1/2$, and (d)--(f) in the radial–axial plane at $\theta = 0$, all for $\sigma=1$. The displacements are induced by a point force acting in the $z$-direction. Remaining parameters are identical to those used in Fig.~\ref{fig:displacement_field_R_The_plane} for $\sigma = 1$. Here, the depicted scaled displacement field is defined as $\bm{U}^* = ( \bm{u} - \bm{u}^\infty ) / ( F_z / (16\pi\mu\rho) )$. It therefore shows the contributions to the displacement field solely induced by the presence of the boundaries, that is, we subtract the solution for a three-dimensionally infinitely extended domain. Considered boundary conditions are (a,d) 2NOS, (b,e) 2FRS, and (c,f) NOS–FRS.
}
    \label{fig:displacement_field_image}
\end{figure*}

In Fig.~\ref{fig:displacement_field_R_The_plane}, we present an exemplary scaled displacement field in the radial-azimuthal plane at $z/\rho=1/2$ for a semi-angle of the wedge of $\alpha = \pi/6$ and a singularity positioned at the polar angle $\beta = \alpha/2$. The material is assumed to be linearly elastic with $\sigma = 1$ (incompressible) for the top row and $\sigma = 2$ $(\nu = 1/4)$ for the bottom row.
Results are shown for three types of boundary conditions: 2NOS in (a) and (d), 2FRS in (b) and (e), and NOS--FRS in (c) and (f).
Numerical integration is performed using $N = 100$ collocation points, providing a well-resolved displacement field.  
In the 2FRS and NOS--FRS cases, the largest magnitudes of displacement are observed near polar azimuthal coordinates of the singularity, $(r,\theta) \simeq(\rho,\beta)$. Contrarily, for the 2NOS case, the region of larger displacement is shifted toward $\theta \simeq 0$. Apparently, the no-slip surface condition enforced on both boundaries shifts the location of the maximum displacement away from the boundary of the wedge towards the center.
Overall, the case $\sigma = 1$ results in magnitudes of displacement approximately twice as large as those for $\sigma = 2$. For incompressible materials ($\sigma=1$), imposed displacements cannot be accommodated by local compressions or dilations. Given that $\sigma = 1$ also formally corresponds to the situation of low-Reynolds-number hydrodynamics of incompressible fluids, these results are particularly useful for validating numerical methods. Specific examples include those based on the boundary element method with a meshed wedge-like boundary~\cite{shum2015hydrodynamic}.

Figure~\ref{fig:displacement_field_R_Z_plane} presents the corresponding results in the radial-axial plane using the same set of parameters as in Fig.~\ref{fig:displacement_field_R_The_plane}. Here, the plots are shown in the azimuthal plane of \( \theta = 0 \).  
We observe that the pattern induced by a point force exhibits similarities between 2NOS and NOS--FRS conditions. However, the overall magnitude of the displacement field for the 2NOS condition is lower than under NOS--FRS conditions.
Illustratively, no-slip conditions are stricter and additionally constrain the displacement field.
In the 2NOS case, pronounced vortex structures are observed, particularly for $\sigma = 1$.
These vortices represent closed lines of displacement that arise due to the boundary conditions, significantly influencing the local appearance. 
Remarkably, the displacement field in this case exhibits two distinct maxima symmetrically located about the plane $z=0$. As $\sigma$ is increased to $2$, these maxima shift toward $z=0$  and merge.

In Fig.~\ref{fig:displacement_field_image}, we present that part of the solution required to satisfy the imposed boundary conditions, using the same parameters as in Fig.~\ref{fig:displacement_field_R_The_plane} for $\sigma = 1$. That is, we subtract from the solution $\bm{u}$ the displacement field $\bm{u}^\infty$ for a three-dimensionally infinitely extended system. The results are shown (a)--(c) in scaled form in the radial–azimuthal plane at $z/\rho = 1/2$, and (d)--(f) in the radial–axial plane at $\theta = 0$.
In the radial–azimuthal plane in (a)–(c), the displacement field is oriented away from the boundaries of the wedge. When combined with the infinite-space solution in an unbounded elastic medium, this ensures that the overall field satisfies the boundary conditions.
In the radial–axial plane in (d)–(f), the displacement field descends for 2NOS and NOS–FRS, while for 2FRS it ascends, following the direction of the applied point force. This behavior results from enforcing no-slip boundary conditions at one or both boundaries.
We mention that in the context of low-Reynolds-number hydrodynamics ($\sigma=1$) it is well known that boundary-induced disturbances of the flow can cause migration toward or away from boundaries for polymers \cite{ma2005theory}, non-spherical particles \cite{mitchell2015sedimentation}, or self-propelled active particles and microswimmers \cite{berke08, spagnolie12, daddi2019frequency}.

Overall, the method introduced in this paper enables an accurate description of the displacement field and precise quantification of its magnitude without the need for computationally extensive fully three-dimensional numerical solutions or simulations, such as those based on finite-element approaches.
Our method allows to calculate the displacement field through a simple one-dimensional numerical integration, which is easy to implement on a basic computer and executes efficiently.

\section{Solution in the limit of a planar boundary}
\label{sec:greens_function_planar_boundary}

Finally, we attempt to recover known results for the specific case of a planar boundary by taking the limit of the semi-opening angle of the wedge $\alpha \to \pi/2$. Depending on the boundary conditions, the expressions for~$\psi_z$ and~$\psi_j$, with $i \in \{x, y\}$, as defined by Eqs.~\eqref{eq:FZ} and~\eqref{eq:FXFY}, respectively, can be derived. 

For 2NOS boundary conditions, we obtain
\begin{subequations}
    \begin{align}
    \psi_z &= q_\parallel \ch \left( (\theta+\beta){p} \right) \, , \\
    \psi_x &= -\frac{q_\parallel \cos\beta}{2\sigma} \, 
    \ch \left( (\theta+\beta){p} \right) \, ,
\end{align}
\end{subequations}
and $\psi_y = 0$. In the case of 2FRS conditions, we arrive at
\begin{equation}
    \psi_z = -q_\parallel \ch \left( (\theta+\beta){p} \right) \, ,
\end{equation}
together with $\psi_x = \psi_y = 0$, as is the case for general $\alpha$ as well.

Thus, the solution for 2NOS conditions in the planar boundary limit can be expressed as
\begin{subequations}
    \begin{align}
    \phi_z &= \frac{q_\parallel}{\left( \rho r\right)^\frac{1}{2}} \, 
    \mathscr{L}_1(\xi, \theta+\beta) \, , \\
    \phi_x &= -\frac{q_\parallel \cos\beta}{\sigma} \, 
    z M^\frac{1}{2} \, 
    \mathscr{L}_2 (\xi, \theta+\beta) \, ,
\end{align} 
\end{subequations}
where
\begin{equation}
    M = \frac{\rho}{r \left( z^2+\left(r-\rho \right)^2\right)
    \left( z^2+\left(r+\rho \right)^2\right)
    } \, .
\end{equation}
Here, we have defined, for convenience, the following two improper integrals,
\begin{subequations} \label{eq:2ints}
\begin{align}
    \mathscr{L}_1 (\xi, a) &= \int_0^\infty \frac{\ch(a{p})}{\ch(\pi{p})} \, 
    P_{i{p}-\frac{1}{2}}(\xi) \, \mathrm{d}{p} \, , \\[5pt]    
    \mathscr{L}_2 (\xi, a) &= 
    \int_0^\infty \left( 4{p}^2+1 \right) \frac{\ch(a{p})}{\ch(\pi{p})} \, 
    P_{i{p} - \frac{1}{2}}^{-1}(\xi) \, \mathrm{d}{p} \, ,
\end{align}
\end{subequations}
where $a = \theta+\beta \in [-\pi,\pi]$.
Analytical expressions for these integrals are provided in Appendix~\ref{appendix:planarboundarylimit}.
For 2FRS conditions, we obtain
\begin{equation}
     \phi_z = -\frac{q_\parallel}{\left( \rho r\right)^\frac{1}{2}} \, 
    \mathscr{L}_1(\xi, \theta+\beta) \, 
\end{equation}
and $\phi_x = 0$.

From there, we obtain after simplification for the 2NOS conditions
\begin{equation}
    \phi_z = \frac{q_\parallel}{\overline{s}} , \quad
    \phi_x = -\frac{2q_\parallel \rho z \cos\beta}{\sigma \overline{s}^3} \, ,  
\end{equation}
and for the 2FRS conditions
\begin{equation}
    \phi_z = -\frac{q_\parallel}{\overline{s}} , \quad
    \phi_x = 0 \, .
\end{equation}
Here, 
\begin{equation}
    \overline{s} = 
    \left( r^2+\rho^2+2\rho r \cos(\theta+\beta)+z^2 \right)^\frac{1}{2} \, ,
\end{equation}
represents the distance from the image of the point force relative to the planar interface.

The corresponding expressions for the image displacement fields are given by
    \begin{align}
    U_r &= q_\parallel\, \frac{z}{\overline{s}^3}
    \left( \rho\cos(\theta-\beta)-r + w \left(r+\rho \cos(\theta+\beta) \right) \right) \, \notag \\
    U_\theta &= -q_\parallel \, \frac{\rho z}{\overline{s}^3}
    \left( \sin(\theta-\beta) + w \sin(\theta+\beta) \right) , \notag \\
    U_z &= \frac{q_\parallel}{\overline{s}} \left( (w-1) \, \frac{z^2}{\overline{s}^2} -\sigma - \frac{w}{3} \right) , \notag
\end{align}
for 2NOS boundary conditions, where we have defined the abbreviation
\begin{equation}
    w = \frac{6\rho r \cos\beta \cos\theta}{\sigma \overline{s}^2} \, . 
\end{equation}
For 2FRS boundary conditions, we obtain
    \begin{align}
    U_r &= q_\parallel \, \frac{z}{\overline{s}^3}
    \left( r + \rho \cos(\theta+\beta) \right) \, , \notag \\
    U_\theta &= -q_\parallel \, \frac{\rho z}{\overline{s}^3} \, \sin(\theta+\beta) \, , \notag \\
    U_z &= \frac{q_\parallel}{\overline{s}}
    \left( \sigma + \frac{z^2}{\overline{s}^2} \right) \, . \notag
\end{align}
These results agree with previous ones for both no-slip and free-slip boundary conditions~\cite{menzel2017force}.

\section{Conclusions}
\label{sec:conclusions}

In summary, we consider linearly elastic, isotropic, spatially homogeneous, possibly compressible media confined to a wedge-shaped geometry. That is, they are bounded by two flat surfaces that meet at a straight edge. No-slip and free-slip surface conditions are applied. We derive expressions for the resulting Green's function under localized application of a constant force that is oriented parallel to the edge. This Green's function quantifies the resulting elastic displacements in the material in response to the force. If we set the Poisson ratio of the elastic material equal to one half, interpret the shear modulus as the shear viscosity and the displacement field as the flow field, then our expressions for appropriate surface conditions equally well apply to the flow of an incompressible viscous fluid under low-Reynolds-number conditions. 

Concerning possible applications of our expressions, we may think of any overdamped motion in a viscous fluid along an edge of the confining vessel. When we combine two point forces to a force dipole, then the motion of active microswimmers along edges could be described to lowest order in the force distribution. Corresponding experiments have been reported \cite{das2015boundaries}. 
Moreover, the slipping of soft gels along the edge of a vessel can be quantified, if driven by forces on discrete inclusions directed along the edge. An example is given by magnetic elastomers and gels that contain magnetic particles in a soft elastic carrier medium \cite{huang2016buckling, puljiz2016forces}. The magnetic particles can be addressed by external magnetic fields, or mutual interactions between them are induced. Such materials may serve as soft magnetic actuators. 

Beyond the examples already mentioned, the derived Green’s function for a wedge-shaped geometry opens up a range of further applications. In microfluidic systems, wedge-like features commonly appear in channel intersections or patterned substrates; here, the Green’s function can aid in modeling electrokinetic flows under applied electric fields. Similarly, in acoustofluidic contexts, it may help describe acoustic streaming effects and wave propagation in confined viscous media. The hydrodynamic interactions of colloidal particles or microswimmers near wedge boundaries can also be quantified, shedding light on their aggregation or diffusive behaviors. 
The Green's function further supports the design of edge-guided actuation in soft robotics, where embedded stimuli in a wedge-like structure can direct movement or amplify local responses. Finally, in biological systems, many tissues feature sharp anatomical boundaries—such as near joints or organ interfaces—and the Green’s function may contribute to modeling localized deformations due to internal or external forces in these complex geometries. 

We remark that the geometry of the wedge-shaped geometry itself has no intrinsic length scale. This makes our solution rather general in application, as long as our assumed conditions such as linear elasticity and low-Reynolds-number flows are met. The relative position of applying the force within the wedge enters the expression of the Green's function.

\section*{Data availability}

Mathematica scripts used to generate the data files are included in the Supporting Information.

\section*{Author contribution}

A.D.M.I.\ and A.M.M.\ designed the research and wrote the manuscript. A.D.M.I.\ performed the analytical calculations and created the figures. 
A.D.M.I., L.F., M.P., and A.M.M.\ reviewed, edited, and provided feedback on the manuscript.

\begin{acknowledgments}
L.F.\ and A.M.M.\ thank the Deutsche For\-schungs\-ge\-mein\-schaft (German Research Foundation, DFG) for support through the Research Unit FOR 5599, grant nos.\ ME 3571/10-1 and ME 3571/11-1. Moreover, A.M.M.\ acknowledges support by the DFG through the Heisenberg Program, grant no.\ ME 3571/4-1. 
\end{acknowledgments}


\appendix 

\section*{Appendices}

The Appendices provide additional technical details on certain results and formulas presented in the main body of the paper without proof.  
Appendix~\ref{appendix:FKL_properties} summarizes key properties of the FKL transform relevant to the current context.  
Appendix~\ref{appendix:derivatives} presents the axial and radial derivatives of the kernel functions $\mathcal{K}_{i{p}}$ and $\mathcal{Q}_{i{p}}$, which are essential for computing the displacement field.  
In Appendix~\ref{appendix:planarboundarylimit}, we derive analytical expressions for the improper integrals $\mathscr{L}_1$ and $\mathscr{L}_2$, which are crucial for determining the displacement field for 2NOS and 2FRS boundary conditions in the planar boundary limit.

\section{Properties of FKL}
\label{appendix:FKL_properties}

In the following, we provide the FKL transform of the Laplace operator, as well as the transform of $(z/r) f(r,z)$, as stated in Eq.~\eqref{eq:FKL_rzf} of the main body of the paper. 
To establish the desired identities, we first determine a few preliminary transforms.

\subsection{Transform of \texorpdfstring{$\mathscr{T}_{i{p}} \left\{ \partial_z f \right\}$}{void}}

This property, identical to that of the Fourier transform, is derived using integration by parts as
\begin{equation}
    \mathscr{T}_{i{p}} \left\{ \frac{\partial f}{\partial z} \right\} = -ik \mathscr{F}_{i{p}} \{f\} \, . \label{eq:app.diff_f_z}
\end{equation}

It also follows that differentiating twice with respect to~$z$ gives
\begin{equation}
   \mathscr{T}_{i{p}} \left\{ \frac{\partial^2 f}{\partial z^2} \right\} = -k^2 \mathscr{F}_{i{p}} \{f\} \, . \label{eq:app.diff_f_zz}
\end{equation}


\subsection{Transform of \texorpdfstring{$\mathscr{T}_{i{p}} \left\{ f/r \right\}$}{void}}

Using the recurrence relation that connects a modified Bessel function of the second kind to the previous and next orders, namely
\begin{equation}
    \frac{K_{i{p}}(|k|r)}{r} = 
    \frac{|k|}{2i{p}} \left(
    K_{i{p}+1}(|k|r) - K_{i{p}-1}(|k|r) \right) ,
\end{equation}
we arrive at
\begin{equation}
\mathscr{T}_{i{p}} \left\{ \frac{f}{r} \right\} = 
\mathcal{C}_1^+ + \mathcal{C}_1^- \, , 
\label{eq:app.f_over_r}
\end{equation}
where 
\begin{equation}
    \mathcal{C}_1^\pm = 
    \pm \frac{|k|}{2i{p}} \,\mathscr{T}_{i{p}\pm 1} \{f\} \, .
\end{equation}


\subsection{Transform of \texorpdfstring{$\mathscr{T}_{i{p}} \left\{ (z/r) \, f \right\}$}{void}}

By differentiating with respect to \( k \) the FKL transform of \( f/r \), we obtain
\begin{align}
ik \mathscr{T}_{i{p}} \left\{ \frac{z}{r} \, f\right\}
= 
    \left( k\, \frac{\partial}{\partial k} 
    - i{p} \right) \mathscr{T}_{i{p}} \left\{ \frac{f}{r} \right\}
    +|k| \mathscr{T}_{i{p}+1} \{ f\} . \notag 
\end{align}
Using Eq.~\eqref{eq:app.f_over_r}, which provides the FKL transform of \( f/r \), we find
\begin{equation}
    \mathscr{T}_{i{p}} \left\{ \frac{z}{r} \, f \right\} 
    =  \mathcal{C}_2^+ 
    + \mathcal{C}_2^-  , \label{eq:app.z_over_r}
\end{equation}
with 
\begin{equation}
    \mathcal{C}_2^\pm = 
    -\frac{\sgn k}{2{p}} \left(
    i{p} \pm 1 \pm k\, \frac{\partial}{\partial k} \right) \mathscr{T}_{i{p}\pm 1} \left\{ f\right\}  ,
\end{equation}
where $\sgn x = x/|x|$ denotes the sign function.


\subsection{Transform of the Laplacian}

The Laplace equation takes a particularly simple form when using the FKL transform. It is more straightforward to compute the FKL transform of $r^2 \Delta f$. We refer to Eq.~\eqref{eq:Laplcian} for the expression of the Laplace operator in cylindrical coordinates. For the radial part of the Laplacian, we apply integration by parts twice. 
In each case, the boundary terms resulting from partial integration vanish upon application of the boundary conditions.
We leverage the fact that
\begin{equation}
    \frac{\partial}{\partial r}
    \left( r\, \frac{\partial}{\partial r} K_{i{p}} (|k|r) \right)
    = \frac{1}{r}\left( k^2 r^2-{p}^2 \right) K_{i{p}} (|k|r) \, 
\end{equation}
along with Eq.~\eqref{eq:app.diff_f_zz}, to obtain
\begin{equation}
    \mathscr{T}_{i{p}} \left\{ 
    r\, \frac{\partial f}{\partial r}
    + r^2 \left( \frac{\partial^2 f}{\partial r^2} + \frac{\partial^2 f}{\partial z^2} \right) \right\}
    = -{p}^2 \mathscr{T}_{i{p}} \{f\} \, .
    \label{eq:app.Lap_axisymm}
\end{equation}
This final results reads
\begin{equation}
    \mathscr{T}_{i{p}} \left\{ r^2 \Delta f \right\}
    = \left( \frac{\partial^2}{\partial\theta^2}-{p}^2 \right) \mathscr{T}_{i{p}} \{f\} \, .
\end{equation}


\subsection{A few remarks on parity with respect to~$z$}

Finally, we offer a few remarks on parity considerations with respect to~$z$.
Both the Fourier and FKL transforms yield a real quantity for an even function of~\( z \) and a purely imaginary quantity for an odd function of~\( z \). Consequently, for symmetric functions that are either even or odd with respect to~$z$, the FKL transforms impose specific relationships between the coefficients \( \mathcal{C}_j^\pm \).

If \( f \) is an even function of \( z \), then for transforms that conserve parity, as is the case for $\mathscr{T}_{i{p}} \{f/r\}$, the result of the transform is a real number. Consequently, the coefficients \( \mathcal{C}_1^\pm \) are complex conjugates, leading to  
\begin{equation}  
    \mathcal{C}_1^+ + \mathcal{C}_1^- = 2 \operatorname{Re} \left\{ \mathcal{C}_1^\pm \right\} \, . 
\end{equation}

If \( f \) is an even function of \( z \), then for transforms that reverse parity in $z$, as in the case of $\mathscr{T}_{i{p}} \{(z/r)f\}$, the result of the transform is a purely imaginary number. As a result, the coefficients \( \mathcal{C}_2^\pm \) are negative complex conjugates, meaning they share the same imaginary part while their real parts have opposite signs. Therefore,  
\begin{equation}  
    \mathcal{C}_2^+ + \mathcal{C}_2^- = 2i \operatorname{Im} \left\{ \mathcal{C}_2^\pm \right\} \, .
\end{equation}  

A similar analysis applies when~\( f \) is an odd function of~\( z \), leading to the opposite outcome.
Specifically, \(\mathcal{C}_1^\pm\) are negative mutual complex conjugates, whereas \(\mathcal{C}_2^\pm\) are complex conjugate pairs.
Note that in the current problem, the functions undergoing this transformation, namely $\chi$ and $\phi_z$, are even in $z$.

For instance, in Ref.~\onlinecite{sano1978effect}, the FKL transform of $(z/r) f$ is expressed in terms of the imaginary part when presenting the coefficients $C^A$ and $C^S$ in their Eq.~(3.9), where~$f$ is an even function of~$z$. In contrast to that, the general formulation in Eq.~\eqref{eq:app.z_over_r} yields significantly simpler expressions, which have proven highly useful in our calculations.

\section{Derivatives of \texorpdfstring{$\mathcal{K}_{i{p}}$}{void} and \texorpdfstring{$\mathcal{Q}_{i{p}}$}{void} with respect to \texorpdfstring{$r$}{void} and \texorpdfstring{$z$}{void}}
\label{appendix:derivatives}

In this Appendix, we provide the expressions for the radial and axial derivatives of \( \mathcal{K}_{i{p}}(r, z) \) and \( \mathcal{Q}_{i{p}}(r, z) \), as defined by Eqs.~\eqref{eq:Kinu} and \eqref{eq:Qnui}, respectively.
These expressions are necessary for the determination of the displacement field, see Eqs.~\eqref{eq:displacement_field_KQ}.

The radial derivative of $\mathcal{K}_{i{p}}$ is obtained as
\begin{equation}
    \frac{\partial \mathcal{K}_{i{p}}}{\partial r} = 
    \frac{ c_- P_{i{p} - \frac{1}{2}}(\xi) + c_+ P_{i{p} + \frac{1}{2}} (\xi) }{2(\rho r)^\frac{3}{2} \left(\xi^2-1 \right) \ch (\pi{p})} ,
\end{equation}
where
\begin{subequations}
    \begin{align}
    c_- &= \rho-\xi r + 2i \xi \left( \xi\rho-r \right) {p} \, , \\
    c_+ &= (r-\xi\rho) (1+2i{p}) \, .
\end{align}
\end{subequations}
The axial derivative of \( \mathcal{K}_{i{p}} \) is straightforward, as it is related to \( \mathcal{Q}_{i{p}} \), see Eqs.~\eqref{eq:Kinu_def} and~\eqref{eq:Qinu_def} for the corresponding definitions of the integrals. 
Specifically, 
\begin{equation}
\frac{\partial \mathcal{K}_{i{p}}}{\partial z} = -\frac{\mathcal{Q}_{i{p}}}{4\rho} \, .
\end{equation}

For $\mathcal{Q}_{i{p}}$, we obtain
\begin{equation}
    \frac{\partial \mathcal{Q}_{i{p}}}{\partial r} =
    \frac{z}{2r}  
    \frac{ d_- P_{i{p} - \frac{1}{2}}^{-1} (\xi) + d_+ P_{i{p} + \frac{1}{2}}^{-1} (\xi) }
    {(\rho r)^\frac{3}{2} \left(\xi^2-1 \right)^\frac{3}{2} } 
    \frac{4{p}^2+1}{\ch(\pi{p})} \,
    ,
\end{equation}
where
\begin{subequations}
    \begin{align}
    d_- &= 3\left(\rho-\xi r\right) + 2i \xi \left( \xi\rho-r \right) {p} \, , \\
    d_+ &= (r-\xi\rho) (3+2i{p}) \, .
\end{align}
\end{subequations}
Moreover,
\begin{equation}
    \frac{\partial \mathcal{Q}_{i{p}}}{\partial z} = 
    \frac{ e_- P_{i{p} - \frac{1}{2}}^{-1} (\xi) + e_+ P_{i{p} + \frac{1}{2}}^{-1} (\xi) }
    {2r (\rho r)^\frac{3}{2} \left(\xi^2-1 \right)^\frac{3}{2} } 
    \frac{4{p}^2+1}{\ch(\pi{p})} \,
    ,
\end{equation}
where
\begin{subequations}
    \begin{align}
    e_- &= 2r\rho \left( \xi^2-1\right) - \xi z^2 \left( 3+2i{p}\right) \, , \\
    e_+ &= z^2 \left( 3+2i{p}\right) \, .
\end{align}
\end{subequations}


\section{Analytical evaluation of improper integrals in the planar boundary limit}
\label{appendix:planarboundarylimit}

Finally, we provide exact analytical expressions for the two indefinite integrals defined by Eqs.~\eqref{eq:2ints}, obtained in the limit of a planar boundary. Since these integrals cannot be evaluated using current computer algebra systems, we perform the calculations manually to obtain the corresponding expressions.

We utilize the integral representation of the Legendre and associated Legendre functions,
\begin{equation}
    P_{q}^m (z) = \frac{(-{q})_m}{\pi}
    \int_{0}^\pi \frac{\cos(mt) \, \mathrm{d}t}{\left( z + \left( z^2-1\right)^\frac{1}{2} \cos t\right)^{{q}+1}} \, , 
    \label{eq:LegendreIntRep}
\end{equation}
with $(z)_n = \Gamma(z+n)/\Gamma(z)$ denoting the Pochhammer symbol~\cite{abramowitz72}. Clearly, the unassociated Legendre function corresponds to the case $m=0$.
Note that $\xi \ge 1$.

\subsection{Integral \texorpdfstring{$\mathscr{L}_1$}{void}}

First, we consider integrals of the form involving the Legendre function
\begin{equation}
    \mathscr{L}_1 (\xi, a) = \int_0^\infty \frac{\ch(a{p})}{\ch(\pi{p})} \, 
    P_{i{p}-\frac{1}{2}}(\xi) \, \mathrm{d}{p} \, ,
\end{equation}
where $a \in[-\pi,\pi]$.
Using Eq.~\eqref{eq:LegendreIntRep} and interchanging the order of integration, we express $\mathscr{L}_1$ as
\begin{equation}
    \mathscr{L}_1 = \frac{1}{\pi} \int_0^\pi
    \mathrm{d}t \int_0^\infty
    \frac{\ch(a{p})}{\ch(\pi{p})} \, \frac{\mathrm{d}{p}}{\left( \zeta(t) \right)^{i{p}+\frac{1}{2}}} \, , 
\end{equation}
with
\begin{equation}
    \zeta(t) = \xi + \left( \xi^2-1\right)^\frac{1}{2} \cos t \, . 
\end{equation}
Integration with respect to \( p \) is performed using the residue method~\cite{brown2009complex} by integrating along a rectangular contour in the \emph{lower} half-plane. Specifically, we choose an clockwise-oriented contour in the complex plane, composed of four rectilinear segments. The vertices are positioned at \( (-R,0) \), \( (R,0) \), \( (R,-R) \), and \( (-R,-R) \).  
In the limit \( R \to \infty \), only the integral along the real axis does not vanish, and it evaluates to \( 2i\pi \) times the sum of the residues at \( z = -(2n+1)i/2 \) for \( n \geq 0 \).
We obtain
\begin{equation}
    \mathscr{L}_1 = \frac{1}{\pi} \int_0^\pi
    \sum_{n=0}^\infty \frac{(-1)^n}{\left( \zeta(t) \right)^{n+1}} \, \cos \left( \left( n+\tfrac{1}{2}\right) a \right) \, \mathrm{d}t \, ,
\end{equation}
which simplifies upon evaluating the infinite sum
\begin{equation}
    \mathscr{L}_1 = \frac{1}{\pi} \, \cos \left( \tfrac{a}{2} \right) \int_0^\pi
    \frac{\left( \zeta(t)+1 \right) \, \mathrm{d}t}{\left( \zeta(t) \right)^2+2\, \zeta(t)\cos a +1} \, .
\end{equation}
This integral is more easily evaluated using the substitution
\begin{equation}
    \zeta(t) = \frac{ 1-X \cos\theta }{1+X\cos\theta} \, , \label{eq:zeta}
\end{equation}
with 
\begin{equation}
    X = \left( \frac{\xi-1}{\xi+1} \right)^\frac{1}{2} \, .
\end{equation}
Then, the limits of integration become $[\pi, 0]$.
Accordingly,
\begin{equation}
    \mathrm{d}t = -\left( \frac{2}{\xi+1} \right)^\frac{1}{2} \frac{\mathrm{d} \theta}{1+X\cos\theta} \, .
\end{equation}
We obtain
\begin{equation}
    \mathscr{L}_1 = \frac{ 2^\frac{3}{2}}{\pi} \, \left( \xi+1 \right)^\frac{1}{2} \cos \left( \tfrac{a}{2} \right)
    \int_0^\frac{\pi}{2} \frac{\mathrm{d}\theta}{ Z(\theta) } \, , 
\end{equation}
where
\begin{equation}
    Z(\theta) = (\xi-1)(1-\cos a) \cos^2\theta + (\xi+1)(1+\cos a) \, . \label{eq:Z}
\end{equation}
The latter integral can be evaluated using the substitution $\cos^2\theta = 1/(1+v^2)$, such that $\mathrm{d}\theta = \mathrm{d}v/(1+v^2)$, resulting in
\begin{equation}
    \mathscr{L}_1 = \frac{1}{2} \left( \frac{2}{\xi + \cos a} \right)^\frac{1}{2} \, .
\end{equation}


\subsection{Integral \texorpdfstring{$\mathscr{L}_2$}{void}}

This integral involves the associated Legendre function of order~$-1$ and is related to the evaluation of the polar and azimuthal components of the displacement field. We define
\begin{equation}
    \mathscr{L}_2 (\xi, a) = 
    \int_0^\infty \left( 4{p}^2+1 \right) \frac{\ch(a{p})}{\ch(\pi{p})} \, 
    P_{i{p} - \frac{1}{2}}^{-1}(\xi) \, \mathrm{d}{p} \, ,
\end{equation}
with $a \in [-\pi,\pi]$.
Instead of proceeding as before with the three previous integrals and using Eq.~\eqref{eq:LegendreIntRep}, we follow a shorter and simpler approach. We use some of the results already obtained with the unassociated Legendre functions in terms of the integral representation
\begin{equation}
    P_{q}^m (z) = \frac{(z+1)^\frac{m}{2}(z-1)^\frac{m}{2}}{\Gamma(-m)} 
    \int_1^z \frac{P_{q}(t) \, \mathrm{d}t \,}{\left( t-z\right)^{m+1}}\, .
\end{equation}
For ${q} = i{p} - \frac{1}{2}$ and $m = -1$, this expression reads
\begin{equation}
    P_{i{p}-\frac{1}{2}}^{-1}(\xi) = \left(\xi^2-1 \right)^{-\frac{1}{2}} 
    \int_1^\xi P_{i{p}-\frac{1}{2}} (t) \, \mathrm{d} t \, , 
\end{equation}
for $\xi \ge 1$.
We thus find that
\begin{equation}
    \mathscr{L}_2
    = \left(\xi^2-1 \right)^{-\frac{1}{2}}
    \int_1^\xi \left( 4 \, \frac{\partial^2}{\partial a^2} + 1 \right) \mathscr{L}_1(t,a) \, \mathrm{d} t \, ,
\end{equation}
which is evaluated as
\begin{equation}
    \mathscr{L}_2 = 
    \frac{\left( 2(\xi-1) \right)^\frac{1}{2}}{\xi+1}
    \left( \frac{\xi+\cos a}{\xi+1} \right)^{-\frac{3}{2}}
    \, .
\end{equation}
It is worth noting that $\mathscr{L}_1$ and $\mathscr{L}_2$ diverge for~$a$ outside the range $[-\pi, \pi]$. However, since $a = \theta + \beta$ remains within this interval, convergence is ensured.

\section*{Supporting Information}

The Mathematica files \textit{FS-FS-RThe-Plane}, \textit{NS-NS-RThe-Plane}, and \textit{NS-FS-RThe-Plane} generate output data for the displacement field in the radial-azimuthal plane, corresponding to the 2FRS, 2NOS, and NOS--FRS boundary conditions, respectively. The files \textit{FS-FS-RZ-Plane}, \textit{NS-NS-RZ-Plane}, and \textit{NS-FS-RZ-Plane} provide the corresponding results in the radial-axial plane. The MATLAB scripts \textit{RThe\_Plane\_Plotter} and \textit{RZ\_Plane\_Plotter} read the output from the Mathematica files and plot the resulting streamlines in the radial-azimuthal and radial-axial planes, respectively.

%


\end{document}